\documentclass[12pt,twoside]{article}

\date{} 
\usepackage {graphicx} 
\def\sn2001el{SN\,2001el~}
\def\snm2001ay{SN\,2001ay~}
\def\aap{A\&A\,  }
\def\aj{AJ  }
\def\apj{ApJ\,  }
\def\apjl{ApJ\,  }
\def\apjs{ApJS  }

\def\mnras{MNRAS\,  }

\def\prc{Phys. Rev. C   }
\begin{document} 

\centerline{\bf ADVANCES IN APPLIED PHYSICS , Vol. x, 20xx, no. xx, xxx - xxx} 

\centerline{\bf HIKARI Ltd, \ www.m-hikari.com}

\centerline{\bf http://dx.doi.org/10.12988/}

\centerline{} 

\centerline{} 

\centerline {\Large{\bf 
The light curve in  supernova modeled by
}} 

\centerline{} 

\centerline{\Large{\bf 
a continuous radioactive decay of $^{56}$Ni
  }} 

\centerline{} 

\centerline{\bf {L. Zaninetti}}

\centerline{}

\centerline{Dipartimento  di Fisica ,}

\centerline{Universit\`a degli Studi di Torino,}

\centerline{via P. Giuria 1,  10125 Torino, Italy}

\centerline{}

{\footnotesize Copyright 
$\copyright$ 2013 Zaninetti Lorenzo. 
This is an open access article distributed 
under the Creative Commons Attribution License, 
which permits unrestricted use, 
distribution, and reproduction in any medium, 
provided the original work is properly cited.}

\begin{abstract} 
The UVOIR bolometric light curves are 
usually modeled by the radioactive decay.
In order to model more precisely the 
absolute/apparent magnitude versus 
time relationship  the continuous production 
of radioactive isotopes 
is introduced.
A differential equation of the first order 
with separable variables 
is solved.
\end{abstract}

\section{Introduction}

The production of $^{56}$Ni , see \cite{Kankainen2010},
in the last phase
of the stellar evolution has  been  predicted
by  \cite {Truran1967,Bodansky1968,Matz1990}.
After this theoretical prediction the radioactive
decay was used as an explanation for the observations
of the light curve of supernova (SN),
see  among others \cite{Mazzali1997,Elmhamdi2003,Stritzinger2006,
Krisciunas2011,Okita2012,Chen2013}.
At the same time  the decay  of  $^{56}$Ni produces a
straight line in the absolute/apparent
magnitude versus time relationship of the light curve 
which does not corresponds  to the  observations.
We briefly recall that such a relationship  
presents a concavity.
In order to explain this discrepancy  between theory of
decay and astronomical light  curve  we have developed
a simple model for the continuous
$^{56}$Ni production.
In this  paper  Section \ref{continuous} 
derives and solves the differential equation which 
models the continuous production of  $^{56}$Ni
and Section \ref{astro} shows the application 
of this new model to the light curve  of two  SNs.
\section{The continuous production of radioactive  isotope}
\label{continuous}
The decay of a radioactive  isotope is  modeled by the
following equation
\begin{equation}
-dN = \frac{N}{\tau}
\quad ,
\end {equation}
where $\tau$ is a constant  and the negative sign
indicates that $dN$ is a reduction in the number
of nuclei , see  \cite{Yang2010}.
The integration of this differential equation
of the first order in which the variables can be separated
gives :
\begin{equation}
N(t) = N_0 e^{-\frac{t}{\tau}}
\label{ntradioactive}
\quad ,
\end{equation}
where $N_0$ is the number of nuclei at  $t=0$.
The half life is  $T_{1/2}= ln(2) \; \tau$ .
The absolute magnitude version  of the previous formula
is
\begin{equation}
M = - C \; Log_{10} (N(t)) = {-\frac{t}{\tau}} +k
\quad ,
\label{magnituderadioactive}
\end{equation}
where $M$ is the absolute luminosity, $C$ and $k$ are two constants.
This means that we are waiting  for a straight line
for the absolute magnitude versus time relationship.
The continuous  production of radioactive nuclei 
is modeled by the
following equation
\begin{equation}
-dN = \frac{N}{\tau} dt - P N^{\alpha} dt
\quad ,
\label{eqnproduction}
\end {equation}
where $P$, the production, and $\alpha$, the exponent,
are two adjustable parameters.
In this  differential equation
of the first order the variables can be separated
and the solution is
\begin{equation}
N(t)= \frac{1}{\left( P\tau+{{\rm e}^{{\frac { \left( \alpha-1 \right) t}{\tau}}}}
 \left( {{\it N_0}}^{-\alpha+1}-P\tau \right)  \right) ^
 {\frac{1}{ \alpha-1}}}
\label{ntproduction}
\quad  ,
\end{equation}
where the initial condition $N(0)=N_0$ has  been used.
The absolute magnitude version of the
previous formula is
\begin{eqnarray}
M = - C \; Log_{10} (N(t)) =   &  \nonumber \\
\frac
{
-\ln  (  ( P\tau+{{\rm e}^{{\frac { ( \alpha-1 )
t}{\tau}}}}{{\it N_0}}^{-\alpha+1}-{{\rm e}^{{\frac { ( \alpha-1
 ) t}{\tau}}}}P\tau ) ^{- ( \alpha-1 ) ^{-1}}
 ) +k\ln  ( 2 ) +k\ln  ( 5 )
}
{
\ln  ( 2 ) +\ln  ( 5 )
}
& \; ,
\label{magnitudeproduction}
\end{eqnarray}
where $M$ is the absolute magnitude
and $C$ and $\alpha$ two constants.

\section {Astrophysical applications}
\label{astro}

\begin{figure}
\begin{center}
\includegraphics[width=6cm]{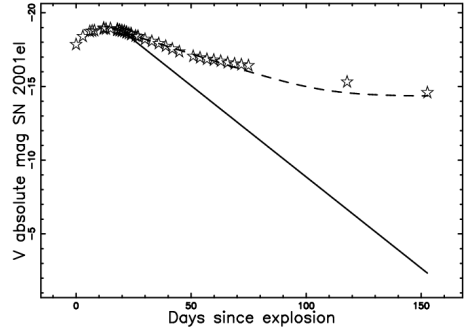}
\end {center}
\caption {
The $V$
light curve of \sn2001el (empty stars)
in absolute magnitude,
the theoretical curve as given
by equation (\ref{magnituderadioactive})
when
the radioactive decay of the  isotope $^{56}$Ni
($\tau $ = 8.757 d  or  $T_{1/2}$ =6.07 d , $k$=-18.65)
was considered
(full line),
and
the theoretical curve
of the
continuous  production of the  isotope $^{56}$Ni
($\tau $ = 8.757 d  or  $T_{1/2}$ =6.07 d , $k$=-18.65,
$P=10^{-4}$, $\alpha$=0.29)
(dashed line).
}
\label{2001elmagvprod}
    \end{figure}

\begin{figure}
\begin{center}
\includegraphics[width=6cm]{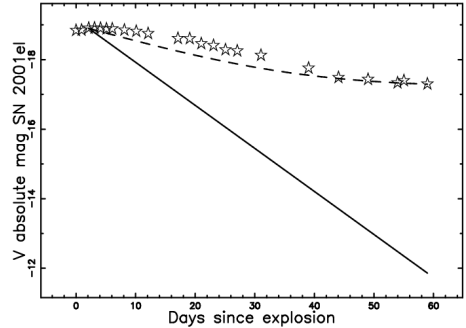}
\end {center}
\caption {
The $V$
light curve of \snm2001ay (empty stars)
in absolute magnitude,
the theoretical curve as given
by equation (\ref{magnituderadioactive})
when
the radioactive decay of the  isotope $^{56}$Ni
($\tau $ = 8.757 d  or  $T_{1/2}$ =6.07 d , $k$=-18.90)
was considered
(full line),
and
the theoretical curve
of the
continuous  production of the  isotope $^{56}$Ni
($\tau $ = 8.757 d  or  $T_{1/2}$ =6.07 d , $k$=-18.90,
$P=0.72 \,10^{-2}$, $\alpha$=0.29)
(dashed line).
}
\label{2001aymagvprod}
    \end{figure}

We  plot  the decay
of the light curve of  \sn2001el , which is of type Ia,
adopting a  distance
modulus of 31.65 mag,
see  \cite{Krisciunas2003},
the  nuclear  decay
which  according to
equation (\ref{magnituderadioactive})
is a straight line,
and the theoretical curve
of the continuous  production of radioactivity
as  represented by equation \ref{magnitudeproduction},
see  Figure \ref{2001elmagvprod}.
Another  example is represented by
\snm2001ay  , the so called
"the most slowly declining type Ia supernova",
which
has distance
modulus of 35.55 mag and is of type Ia, see \cite{Krisciunas2011}.
Figure \ref{2001aymagvprod}
reports  the
light curve, the nuclear decay of the isotope $^{56}$Ni
and  the continuous  production  of the isotope $^{56}$Ni.

\section {Conclusions}

In conclusion
the continuous production of $^{56}$Ni during 
the evolution of a SN 
is here modeled introducing two parameters $\alpha$  and $P$ ,
see eqn.(\ref{eqnproduction}).
The solution of this differential equation of 
the first order with variables
which  can be separated has been derived, 
see eqn.(\ref{ntproduction}). 
The application of this new solution to \sn2001el
and  \snm2001ay produces an acceptable agreement 
between  theory and
observations over the considered temporal interval 
of $\approx$ 60d, see Figs. \ref{2001elmagvprod} and 
\ref{2001aymagvprod}.


\end{document}